\documentclass[aps,prd,reprint,groupedaddress,amsmath,amssymb]{revtex4-2}

\usepackage{graphicx}
\usepackage{bm}
\usepackage{amssymb}
\usepackage{amsmath}
\usepackage{color}
\usepackage{mathtools}
\usepackage{cases}
\usepackage{booktabs}
\usepackage{mathrsfs}
\usepackage{float}
\usepackage{array}
\usepackage{dcolumn}
\usepackage{bm}
\usepackage{braket}

\usepackage[
colorlinks=true,
filecolor=black,
anchorcolor=blue,
linkcolor=blue,
citecolor=cyan, 
urlcolor=cyan,
linktocpage=true,
plainpages=false,
breaklinks=true,
pdfstartview=FitH
]{hyperref}

\begin{document}

\title{Detectability of axion-like dark matter for different time-delay interferometry combinations in space-based gravitational wave detectors}

\author{Yong-Yong Liu $^{1,4,5}$}
\email[co-first author: ]
{liuyongyong23@mails.ucas.ac.cn}
\author{Jing-Rui Zhang $^{1,4,5}$}
\email[co-first author: ]
{zhangjingrui22@mails.ucas.ac.cn}
\author{Ming-Hui Du$^{3}$} 

\author{He-Shan Liu$^{3}$}

\author{Peng Xu$^{3,1,5}$}
\author{Yun-Long Zhang$^{2,1}$}
\email[corresponding author: ]
{zhangyunlong@nao.cas.cn}
\affiliation{$^{1}$School of Fundamental Physics and Mathematical Sciences,  Hangzhou   Institute for Advanced Study, University of Chinese Academy of Sciences, Hangzhou 310024, China}

\affiliation{$^{2}$National Astronomical Observatories, Chinese Academy of Sciences, Beijing, 100101, China}

\affiliation{$^{3}$Center for Gravitational Wave Experiment, National Microgravity Laboratory, Institute of Mechanics, Chinese Academy of Sciences, Beijing 100190, China}

\affiliation{$^{4}$ Institute of Theoretical Physics, Chinese Academy of Sciences, Beijing 100190, China} 

\affiliation{$^{5}$Taiji Laboratory for Gravitational Wave Universe (Beijing/Hangzhou), University of Chinese Academy of Sciences, Beijing 100049, China}

\begin{abstract}
In the space-based gravitational wave detections, the axion-like dark matter would alter the polarization state of the laser link between spacecrafts due to the birefringence effect. However, current designs of space-based laser interferometer  are insensitive to variations in the polarization angle. Thus, the additional wave plates are employed to enable the response of the axion-induced birefringence effect. We calculate and compare the sensitivities of different space-based detectors, accounting for three time-delay interferometry combinations, including Monitor, Beacon, and Relay. We find that the Monitor and Beacon combinations have better sensitivity in the high-frequency range, and the optimal sensitivity reaches $g_{a\gamma}\sim 10^{-13}\text{GeV}^{-1}$, while the Sagnac combination is superior in the low-frequency range. We also find that ASTROD-GW can cover the detection range of axion-like dark matter mass 
down to $10^{-20}$ eV.

\end{abstract}
\maketitle

\tableofcontents
\allowdisplaybreaks

\section{Introduction}\label{Introduction}
The nature of dark matter is one of the greatest mysteries in modern physics. The existence of dark matter has been confirmed by various observations, including galactic rotation curves~\cite{Zwicky:1933gu,Rubin:1970zza,Rubin:1980zd}, the cosmic microwave background~\cite{Planck:2018vyg} and gravitational lensing~\cite{Walsh:1979nx,Clowe:2003tk,Markevitch:2003at}. However, we still do not know its microscopic properties, such as its mass, spin, and possible interactions with ordinary matter. Among various dark matter candidates, axion-like dark matter is one of the most compelling and widely studied theoretical candidates~\cite{Preskill:1982cy,Abbott:1982af,Dine:1982ah,Kim:2008hd,Marsh:2015xka}.

Multiple hypothetical interactions can be used to probe for axion-like dark matter, including its couplings with photons, gluons, spins, and gravitational waves~\cite{Gu__2024,kim2024oscillationsatomicenergylevels, gué2025searchqcdcoupledaxion,annurev:/content/journals/10.1146/annurev-nucl-102014-022120, Gavilan_Martin_2025, miller2025gravitationalwaveprobesparticle}.
Here, we mainly consider the interaction between axion-like dark matter and photons. Based on this interaction, there are mainly two methods of detection. One of them is to utilize the conversion between photons and axion-like dark matter in the presence of a background magnetic field~\cite{PhysRevLett.51.1415}, 
which forms the underlying principle of experiments such as CAST~\cite{CAST:2017uph}. Designed to search for X-rays produced by the conversion of solar axions, CAST has set stringent constraints on the parameter space of axion-photon coupling, and the next-generation experiment IAXO will push the sensitivity beyond current limits~\cite{Armengaud_2019}. Further detection efforts constitute a broad experimental portfolio, including ABRACADABRA-10 cm~\cite{PhysRevLett.122.121802,PhysRevLett.127.081801}, ``light shining through a wall"~\cite{PhysRevD.78.092003,PhysRevD.92.092002}, and other initiatives~\cite{PhysRevD.42.1297,PhysRevLett.104.041301,8395076}. Axion-like dark matter can also be generated in the extremely high-temperature environment within the interior of massive stars during their later stages of evolution, in which a small portion of axion-like dark matter flows out from the stellar atmosphere and decays into photons. These photons may be detected with terrestrial radio telescopes on Earth~\cite{PhysRevD.97.123001,PhysRevLett.121.241102,PhysRevD.99.123021,PhysRevD.101.123003,PhysRevD.102.023504}.  

In addition, with the existence of the background axion field, photons with different chirality would exhibit different phase velocities due to the photon-axion coupling, which is called the axion-induced birefringence effect~\cite{sigl2019axionlikedarkmatterconstraints}.
For linearly polarized light, this phase velocity difference is manifested as the rotation of the polarization plane when passing through the background field. The traditional astrophysical observation methods have been employed to search for the birefringence effect caused by axion-like dark matter, such as the study of cosmic microwave background (CMB) polarization~\cite{Fedderke_2019,HARARI199267,Fujita_2021} and astrophysical polarization measurements~\cite{Alighieri_2010,Liu_2020,Liu_2023,Chen_2022,Yuan__2021}. However, due to system uncertainties and astrophysical foregrounds, the sensitivities for such experiments are limited. Recently, new experimental schemes have been proposed, which utilize high-precision interferometric measurement of tiny phase shifts in optical cavities~\cite{PhysRevD.101.095034,PhysRevLett.102.202001,PhysRevLett.132.191002,pandey2024resultsaxiondarkmatterbirefringent,PhysRevD.100.023548} and laser interferometers~\cite{DeRocco:2018jwe,Nagano:2019rbw,michimura2025searchesultralightvectoraxion}. Among those detection methods, space-based laser interferometers have longer baselines compared to ground-based detectors, eliminating the noise caused by uncontrollable factors on the ground. Therefore, these interferometers may reach high sensitivities for axion-like dark matter at low frequencies.

Recent studies have explored the potential of space-based laser interferometers to detect ultralight dark matter~\cite{Yu:2023iog,Yu:2024enm,Zhang:2025fck}, especially for axion-like dark matter~\cite{Yao_2025,Gu__2025,yao2025axionlikedarkmattersearch}. These studies have adapted existing interferometers to make the detectors sensitive to the axion-induced birefringence effect and have assessed their corresponding sensitivities. However, there are still some gaps in the existing research on axion dark matter detection. For space-based laser interferometers, time delay interferometry (TDI) is essential to suppress dominant noise sources, including laser frequency noise and clock noise. While many different TDI combinations exist, their sensitivities for the detection of axion-like dark matter may vary across different frequency bands. Previous research has primarily focused on a limited set of TDI combinations, and the optimal TDI combination for specific frequency bands still needs to be studied. Here we systematically evaluate the sensitivities of the ASTROD-GW $\mu$Hz-band interferometer for three extra TDI combinations: Monitor, Beacon, and Relay. Through a comparative analysis with LISA-like interferometers utilizing various TDI combinations, we find that Monitor and Beacon combinations provide markedly superior sensitivity at high frequencies. Specifically, they enhance the optimal sensitivity by about an order of magnitude and deliver this sensitivity across a wide frequency band.

The paper is organized as follows. In Sec.~\ref{axion-induced birefringence effect}, we introduce the theoretical framework of axion-induced birefringence and the single-arm response signal of the detector. In Sec.~\ref{Detector and TDI combination}, we introduce the basic information of the ASTROD-GW mission and calculate the one-sided power spectral density (PSD) of the signals for different TDI combinations. In Sec.~\ref{Sensitivity}, We calculate the sensitivity curves of ASTROD-GW, LISA, Taiji and TianQin, and conduct an comparative analysis around various interferometers and different TDI channels. Finally, we conclude in Sec.~\ref{Conclusion}. In this paper, we use natural units $c=\hbar=1$ and the metric signature  as $(-,+,+,+)$.

\section{Axion-induced birefringence effect}
\label{axion-induced birefringence effect}
In this section, we introduce the theoretical formalism of the axion-induced birefringence effect and derive the resulting phase shift for circularly polarized light.
The Lagrangian describing the Chern-Simons interaction between the axion field and the photon field is given by \cite{PhysRevD.41.1231,PhysRevD.43.3789}
\begin{equation}
    \mathcal{L}=-\frac{1}{4}F_{\mu\nu}F^{\mu\nu}-\frac{1}{2}\partial_{\mu}a\partial^{\mu}a-\frac{1}{2}m^{2}a^{2}-\frac{g_{a\gamma}}{4}aF_{\mu\nu}\tilde{F}^{\mu\nu},
\end{equation}
where the last term describes the interaction, and $g_{a\gamma}$ is the corresponding coupling constant. $F_{\mu\nu}$ and $\tilde{F}^{\mu\nu}$ are the electromagnetic field tensor and its dual tensor, respectively.

Since the arm length and the observation time of the interferometer in this study are much shorter than the coherence length and the coherence time of the axion, respectively, we can treat the axion field as a coherent field. Then the axion field can be approximately described by~\cite{Khmelnitsky_2014,Hui_2021,Yao:2024fie}
\begin{equation}
    a(t)=a_0\cos(mt+\theta_0)=a_0e^{i(mt+\theta_0)},
\end{equation}
where $a_{0}$ is the field amplitude that depends on the local dark matter density $\rho_{\text{DM}}$, and $\rho_{\text{DM}}\simeq 0.4~\text{GeV}/\text{cm}^3$. To be specific, the ensemble average satisfies $\langle a_{0}^{2}\rangle=2\rho_{\rm DM}/m^{2}$. The interaction between axion-like dark matter and the photon would modify the dispersion relations of the photon
\begin{equation}
    \omega^2-k^2=\pm g_{a\gamma}\dot{a}k\ ,
\end{equation}
and the modified phase velocities of circularly polarized light are given by
\begin{equation}
    v_{\pm}=\frac{\omega}{k}\simeq 1\pm\frac{g_{a\gamma}\dot{a}}{2k}=1\pm\delta v.
\end{equation}
It should be noted that the axion-induced birefringence effect modifies the phase velocity only for left- and right-handed circularly polarized light, while leaving the phase velocity of linearly polarized light unchanged.

The phase shift caused by the birefringence effect is then given by~\cite{Gu__2025}
\begin{align}
    \Delta\phi(t)&=k\delta L(t)=k\int_{t-L}^{t}\delta v dt'\nonumber\\
    &=\frac{1}{2}g_{a\gamma}[a(t)-a(t-L)]\equiv\phi_{
    \rm single\ link}.
\end{align}
\color{black}
For right-handed circularly polarized light, we define the relative frequency fluctuation signal as
\begin{equation} \label{6}
    \eta_{i}(t)=-\frac{1}{\omega}\frac{\Delta\phi(t)}{dt}=-\frac{im}{2\omega}g_{a\gamma}[a(t)-a(t-L)],
\end{equation}
where $\omega/2\pi=\nu_{0}$ denotes the nominal laser frequency. Similarly, for left-handed circularly polarized light, the signal can be expressed as
\begin{equation} \label{7}
    \eta_{i'}(t)=\frac{1}{\omega}\frac{\Delta\phi(t)}{dt}=\frac{im}{2\omega}g_{a\gamma}[a(t)-a(t-L)].
\end{equation}
Such phase shift induced by the axion-induced birefringence effect can be detected by the interferometer. This provides a promising method to probe axion-like dark matter with laser interferometers~\cite{Yu:2024enm,Yao_2025,Gu__2025,yao2025axionlikedarkmattersearch}.

\section{Laser interferometers
and TDI combinations}
\label{Detector and TDI combination}

In this section, we introduce the major space-based GW detectors operating within frequency band from microhertz to millihertz. We discuss the necessary modifications of the optical path for the interferometers, which are required for the detection of axion-like dark matter. Additionally, we calculate the one-sided power spectral densities of the signals for different TDI combinations.

\begin{figure}[h]
\centering~~~
\includegraphics[keepaspectratio, scale=0.3]{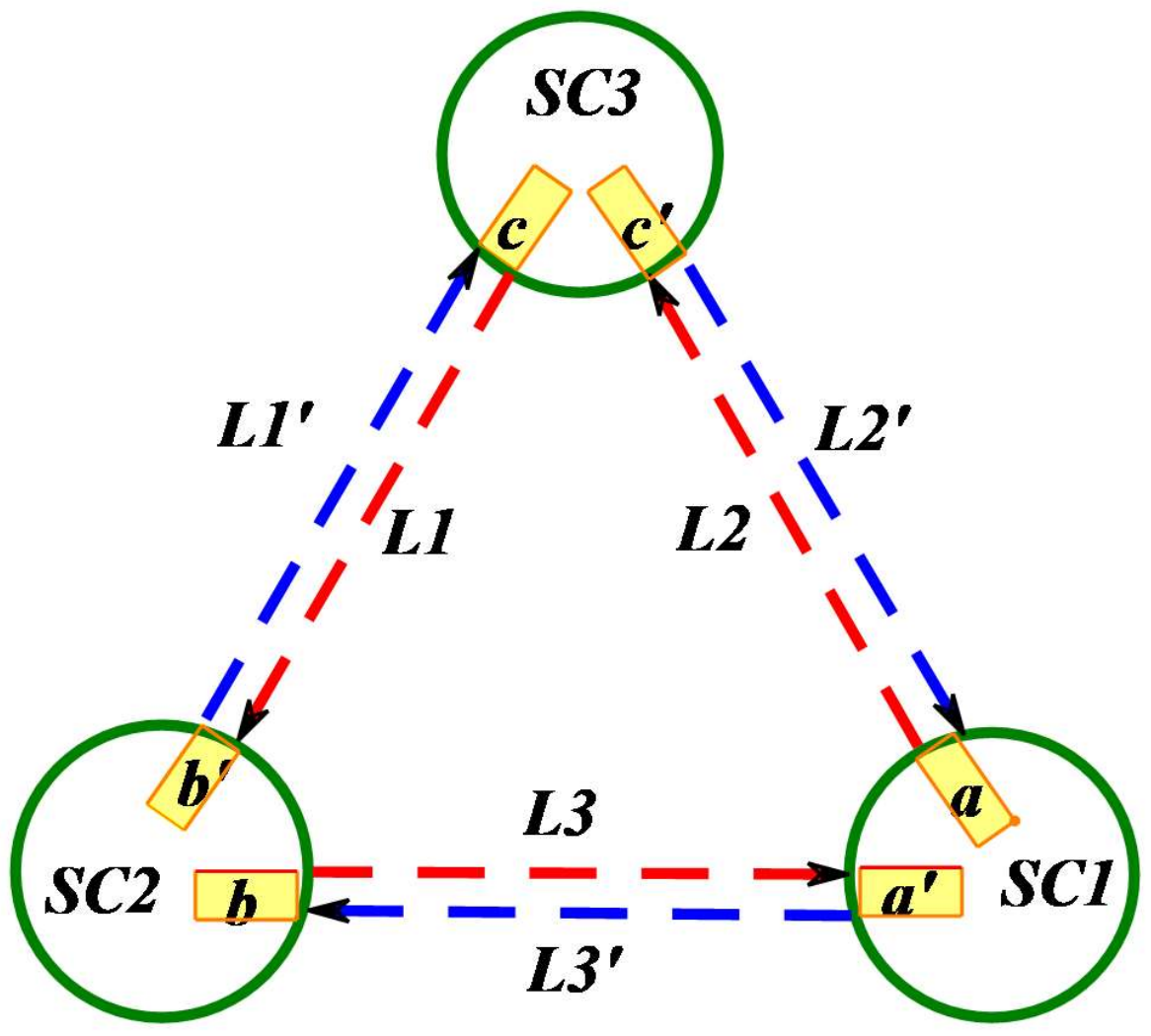}
\caption{Schematic diagram of the configuration of the three spacecrafts in the interferometer. Each spacecraft is equipped with two optical platforms, labeled $j$ and $j'$, where $j\in \{a,b,c\}$ and $j'\in \{a',b',c'\}$. By convention, labels corresponding to light traveling in the clockwise direction are denoted with a prime symbol, for instance, $a'$.}
\label{fig:31}
\end{figure}

\subsection{Space-based gravitational wave detectors}

The basic plan of ASTROD was proposed in 1993 and has since remained in the  conceptual, simulation, and laboratory research phases~\cite{Ni_1996, NI20031437}. In 1996, the Mini-ASTROD mission concept was proposed with the aim of testing general relativity and mapping the gravitational field of the solar system~\cite{Ni_1996ns}. In 2009, ASTROD-GW was proposed for the detection of gravitational waves. The arm length of this mission is designed as $2.6\times 10^{11}\rm m$, and the three spacecrafts operated near the Lagrange points L3, L4 and L5 of the Sun-Earth system, respectively~\cite{Ni_2024acg}. The schematic diagram of the detectors is depicted in Fig.~\ref{fig:31}.

\begin{figure}[h]
    \centering
    \includegraphics[keepaspectratio, scale=0.4]{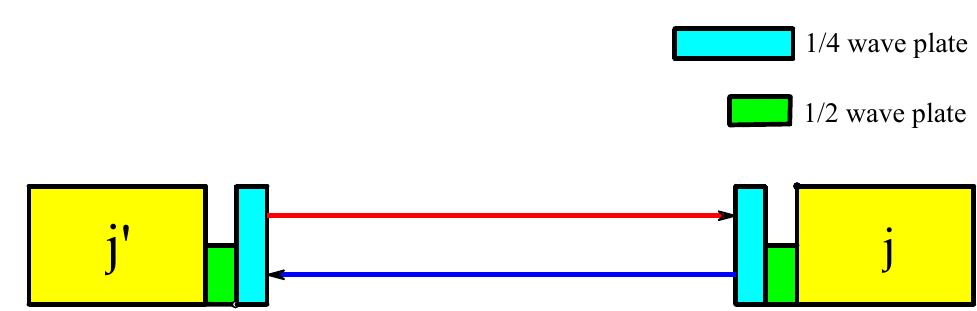}
    \caption{Schematic diagram of the modification to the single laser link. By adding waveplates to each optical path, the emitted and received light is converted into circularly polarized light. The optical path undergoes an alteration such that the light propagating along the clockwise directed optical path(blue line) exhibits left-handed circular polarization, while the light propagating along the counter-clockwise directed optical path(red line) exhibits right-handed circular polarization.}
    \label{fig:32}
\end{figure}

When light propagates through the axion-like field, the birefringence effect gives rise to a periodic variation in its polarization angle. However, conventional interferometer designs are insensitive to such polarization shifts, which necessitates specific modifications. At present, there are two main modification schemes for laser interferometers. One approach entails embedding a dark matter interferometer capable of detecting rotations in the polarization plane~\cite{PhysRevLett.132.191002, yao2025axionlikedarkmattersearch}. Another method involves inserting waveplates into the interferometer, ensuring that both the transmitted and received laser beams are circularly polarized~\cite{PhysRevD.98.035021, Yao_2025, Gu__2025}. This work focuses on the second approach. In Fig.~\ref{fig:32}, we show the modified diagram of a single-link system. By introducing waveplates into the original design, both the transmitted and received light between spacecraft undergo conversion from linear to circular polarization.

After suppressing laser frequency noise and clock frequency noise using TDI technique (detailed in Section~\ref{sec_tdi}), the dominant noise sources limiting sensitivity are test mass acceleration noise and optical metrology system noise. For ASTROD-GW, the one-sided PSD of the test mass acceleration noise $S_{\rm acc}$ and the optical metrology system noise $S_{\rm oms}$ are given by~\cite{Wang:2023jct}
\begin{align}
&S_{\rm acc}=\left(\frac{s_{\rm acc}}{2\pi fc}\right)^2\left[1+\left(\frac{0.1~{\rm mHz}}{f}\right)^2\right]{\rm Hz}^{-1},\label{eq8}\\
&S_{\rm oms}=\left(\frac{2\pi f}{c}s_{\rm oms}\right)^2\left[1+\left(\frac{0.2~{\rm mHz}}{f}\right)^4\right]{\rm Hz}^{-1}.\label{eq9}
\end{align}

\begin{table}[!htp]
	\centering
	\begin{tabular}{l|l|l|l|l}
\hline\hline
parameter & ASTROD-GW & LISA & Taiji & TianQin  \\ \hline
		Arm length $L(10^{9}\rm m)$ & 260  & 2.5 & 3 & 0.17 \\ 
		Frequency\ band(Hz) & $[10^{-7},10^{-1}]$ &$[10^{-5},1]$ & $[10^{-5},1]$ &$[10^{-4},10]$    \\
		$s_{\rm acc}(10^{-15}{\rm m}{\rm /s^{2}})$ & 0.3 & 3  & 3 & 1   \\
		$s_{\rm oms}(10^{-12}{\rm m})$ & 104 & 15 & 8 & 1  \\ \hline\hline
	\end{tabular}
\caption{The specific design parameters of ASTROD-GW~\cite{NI_2013, wang2024timedelayinterferometryastrodgw}, LISA~\cite{amaroseoane2017laserinterferometerspaceantenna}, Taiji~\cite{10.1093/ptep/ptaa083}, and TianQin~\cite{TianQin:2015yph}.}
\label{3bt}
\end{table}

For LISA, Taiji, and TianQin, the power spectral densities of these noises are given by~\cite{babak2021lisasensitivitysnrcalculations}
\begin{align}
&S_{\rm acc}=\left(\frac{s_{\rm acc}}{2\pi fc}\right)^2 \left[1+\left(\frac{0.4~{\rm mHz}}{f}\right)^2\right]\nonumber\\
&\qquad\qquad\qquad \times \left[1+\left(\frac{f}{8~{\rm mHz}}\right)^{4}\right]{\rm Hz}^{-1},\\
&S_{\rm oms}=\left(\frac{2\pi f}{c}s_{\rm oms}\right)^2\left[1+\left(\frac{2~{\rm mHz}}{f}\right)^4\right]{\rm Hz}^{-1}.
\end{align}
The design parameters for different detectors are provided in Table~\ref{3bt}.

\subsection{TDI combinations and signal responses}
\label{sec_tdi}

Time-delay interferometry (TDI) was proposed to address the problem that laser frequency noise in space-based gravitational wave detectors cannot be canceled directly. The core concept involves synthesizing a virtual equal-arm interferometer by time-shifting and linearly combining the measurement data from the unequal-arm links. This method allows for the suppression of laser frequency noise in post-processing while preserving the gravitational wave signal. Since its initial conceptual proposal in the 1990s~\cite{10.1117/12.293329,NI20031437}, TDI has evolved into three generations over the past two decades~\cite{Estabrook:2000ef,Tinto:2004wu}. These correspond to the 
1st-generation combinations for static constellations, the 1.5th-generation for rigidly rotating constellations~\cite{PhysRevD.69.022001}, and the 2nd-generation that accounts for the relative velocities between spacecrafts due to orbital dynamics~\cite{PhysRevD.68.061303}. TDI continues to advance rapidly and can now effectively suppress not only laser frequency noise but also platform jitter and clock noise~\cite{Otto_2012, PhysRevD.98.042003}. Given the similarity in principle between detecting axion-like dark matter signals and gravitational waves with space interferometers, the TDI techniques developed for noise cancellation are directly applicable to axion-like dark matter detection.

\begin{figure}[tbp]
    \centering
    \includegraphics[keepaspectratio, scale=0.64]{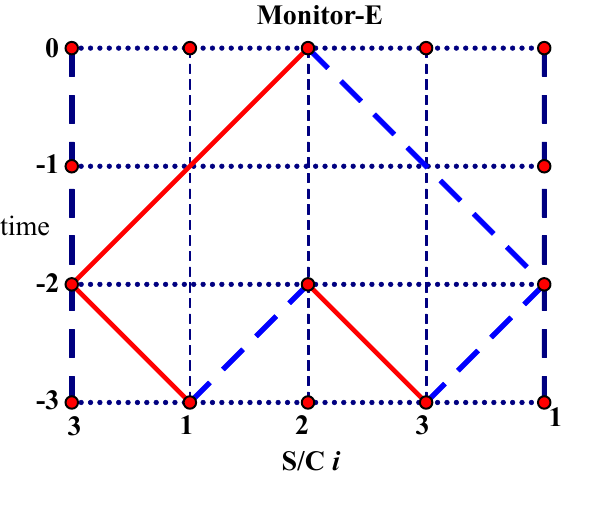}
    \caption{The space-time diagram of Monitor-E combination. The red solid line and the blue dashed line represent two groups of laser links. The horizontal axis indicates the spacecraft's serial number, and the vertical axis represents the time direction.}
    \label{fig:33}
\end{figure}

\begin{figure}[ht]
    \centering
    \includegraphics[keepaspectratio, scale=0.6]{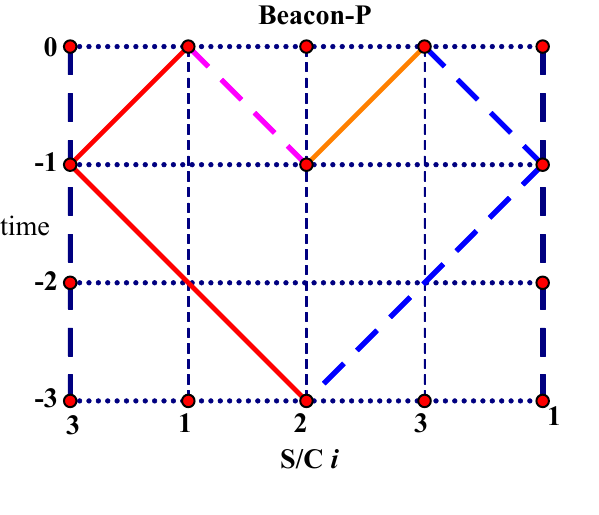}
    \caption{The space-time diagram of the Beacon-P combination. The solid and dashed lines constitute two groups of laser links, with different colors distinguishing the laser links emitted at different times.}
    \label{fig:34}
\end{figure}

\begin{figure}[tbp]
    \centering
    \includegraphics[keepaspectratio, scale=0.6]{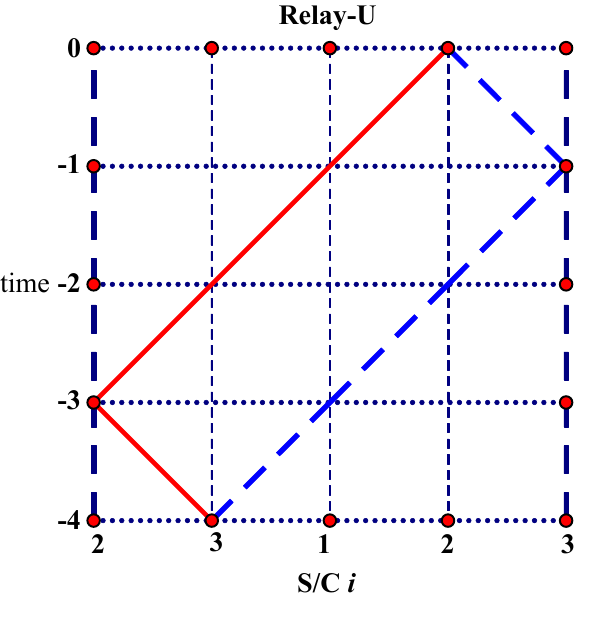}
    \caption{The space-time diagram of the Relay-U combination, where solid and dashed lines represent two groups of laser links. Compared with the Monitor-E and Beacon-P combinations, the Relay-U combination exhibits a longer delay time and a more symmetrical synthesized virtual laser link.}
    \label{fig:35}
\end{figure}

Currently, the most common combinations include Michelson, Sagnac, fully symmetric Sagnac, Beacon, Relay, and Monitor. The first three combinations have already been thoroughly studied in the application of detecting axion-like dark matter, and it has been found that the Sagnac combination can maximize the sensitivity of the interferometer for axion-like dark matter~\cite{Yao_2025,yao2025axionlikedarkmattersearch}. Here we mainly study the remaining three combinations and compare them with the Sagnac combination.


{
The TDI combinations Monitor, Beacon, and Relay consist of $\{E,F,G\}$, $\{P,Q,R\}$ and $\{U,V,W\}$~\cite{PhysRevD.62.042002,Tinto:2020fcc}, respectively. Each group of capital letters represents the signal combination number extracted from different spacecraft vertices in each TDI combination.
Here we mainly consider $E$, $P$ and $U$ combinations, and other combinations can be obtained by permuting the indices.}
We present the space-time diagrams~\cite{wang2024timedelayinterferometryastrodgw,Wang_2021,Wang:2024ckw} for these TDI combinations in Figs.~\ref{fig:33}--\ref{fig:35}. Their expressions are given by~\cite{WANG2012211,PhysRevD.106.044054}
\begin{align}
    E=&\eta_{1}+D_{3}\eta_{2}+D_{31}\eta_{3'}+D_{11'}\eta_{1'}\nonumber\\
    &-(\eta_{1'}+D_{2'}\eta_{3'}+D_{2'1'}\eta_{2}+D_{11'}\eta_{1}),\\
    P=&D_{1}\eta_{1}+D_{13}\eta_{2'}+D_{133'}\eta_{1'}+D_{2'}\eta_{2}\nonumber\\&-(D_{1}\eta_{1'}+D_{2'}\eta_{2'}+D_{2'3'}\eta_{1}+D_{2'33'}\eta_{2}),\\
    U=&-(D_{3'}-D_{11'3'})\eta_{1'}+(1-D_{3'2'1'})\eta_{2}\nonumber\\
    &-(1-D_{11'})\eta_{2'}+(D_{1}-D_{3'2'})\eta_{3'},
\end{align}
where $D_{ijk}\eta_{i}(t)\equiv\eta_{i}(t-L_{i}-L_{j}-L_{k})$. Note that here we employ a linear-combination-based method to combine the signals, instead of subtracting two virtual paths as in the Sagnac combination.

In this study we focus on the first-generation TDI formulation.
For simplicity, we approximate the detector arm lengths to be equal, i.e., $L\equiv L_{1}=L_{2}=L_{3}=L_{1'}=L_{2'}=L_{3'}$. By substituting the single-arm data streams from Eqs.~(\ref{6}) and (\ref{7}) into the three classes of TDI combinations, we obtain their corresponding expressions
\begin{equation}
    C_{i}(t)=\frac{img_{a\gamma}}{\omega}a_{0}e^{i(mt+\theta_{0})}c_{i},
\end{equation}
where $C_{i}\in\{E,P,U\}$. The coefficients for each TDI combination are expressed as:
\begin{align}
    c_{E}&=-1+3e^{-i2mL}-2e^{-i3mL}\nonumber,\\
    c_{P}&=-2e^{-imL}+3e^{-2imL}-e^{-i4mL}\nonumber,\\
    c_{U}&=-1+e^{-imL}+e^{-i3mL}-e^{-i4mL}.
\end{align}
For each TDI combination, the one-sided PSD of the signal can be calculated as
\begin{equation}\label{eq.17}
    P_{s}^{C_{i}}(f)=2\frac{|\tilde{C_{i}}(f)|^{2}}{T}=\frac{\rho_{\text{DM}}g_{a\gamma}^{2}T}{\omega^{2}}|c_{i}|^{2},
\end{equation}
where $T$ is the observation period, the frequency $f$ is related to the mass of axion-like particles via $m=2\pi f$, and $|c_{i}|^{2}$ represent the coefficients for different TDI combinations, given by:
\begin{align}
    |c_{E}|^{2}&=|c_{P}|^{2}=16\left[5+4\cos(mL)\right]\sin^4\left(\frac{mL}{2}\right) \nonumber,\\
    |c_{U}|^{2}&=16\left[1+2\cos(mL)\right]^2\sin^4\left(\frac{mL}{2}\right).
\end{align}
Note that the Monitor and Beacon combinations exhibit identical one-sided PSDs for both signal and noise, as can also be inferred from their space-time diagrams in Figs.~\ref{fig:34}--\ref{fig:35}.

\begin{figure}[h]
    \centering
    \includegraphics[keepaspectratio, scale=0.48]{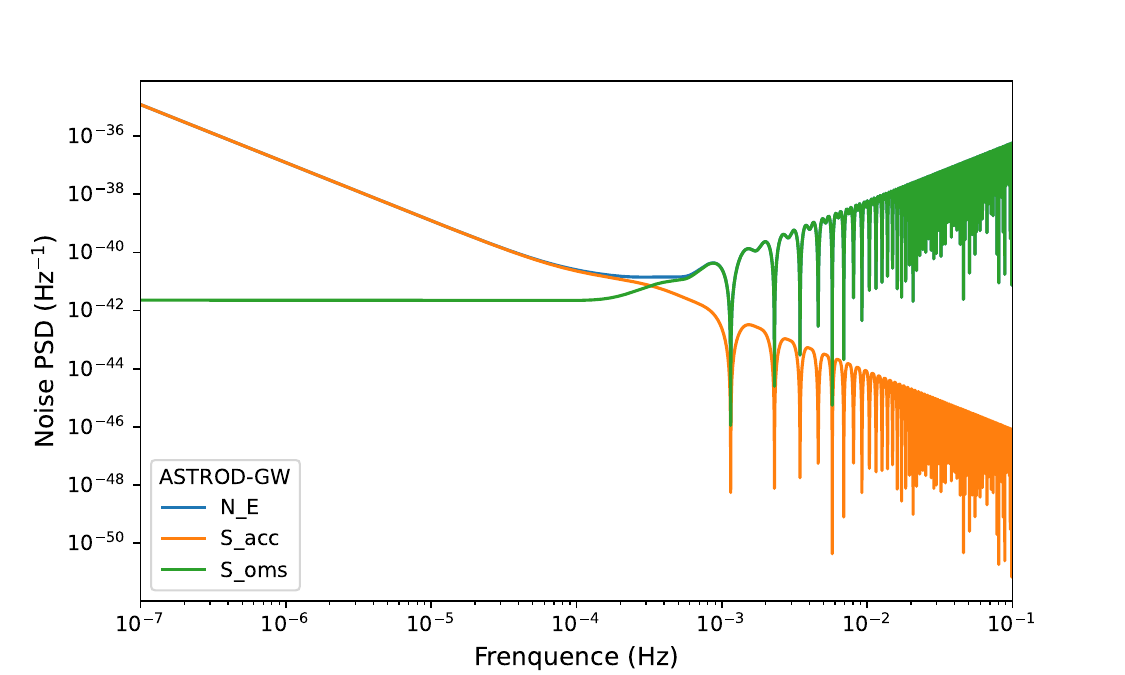}
    \caption{The one-sided noise PSD of the Monitor E combination in ASTROD-GW. The blue line represents the total one-sided noise PSD of the ASTROD-GW Monitor E combination, the orange line represents $S_{\rm acc}$, and the green line represents $S_{\rm oms}$. At low frequencies, the noise is dominated by acceleration noise, while at high frequencies, it is dominated by the optical metrology system noise.}
    \label{47}
\end{figure}

On the other hand, the one-sided PSDs of the noises for these TDI combinations are given by~\cite{Yu:2023iog,Wang:2024hbh,PhysRevD.111.064056,PhysRevD.108.044075}
\begin{align}\label{eq.19}
    P_{n}^{E}(f) &= P_{n}^{P}(f)=16[2\sin^{4}(\pi fL)+\sin^{2}(2\pi fL)]S_{\rm acc}\nonumber\\
    &\qquad \quad +8[\sin^{2}(\pi fL)+\sin^{2}(2\pi fL)]S_{\rm oms},\\
\nonumber
    P_{n}^{U}(f)
&= 8[2\sin^{2}(\pi fL)+\sin^{2}(2\pi fL)+2\sin^{2}(3\pi fL)]S_{\rm acc}\\
&~ +4[\sin^{2}(\pi fL)+2\sin^{2}(2\pi fL)+\sin^{2}(3\pi fL)]S_{\rm oms}.
\label{eq.20}
\end{align}
Here, since the Monitor and Beacon combinations yield identical one-sided PSDs for both signal and noise, their sensitivities are the same. In the following analysis, we consider only the Sagnac-$\alpha$, Monitor-E, and Relay-U combinations.
The one-sided noise PSD of the Monitor E combination in ASTROD-GW is shown in Fig.~\ref{47}.

\section{Sensitivities to the axion-photon coupling}
\label{Sensitivity}

In this section, we define the theoretical expression for the detector sensitivity and present a comparison of the sensitivity curves for the four kinds of detectors under three TDI combinations. We also include the constraints for axion-photon coupling from CAST~\cite{2017} and SN1987A~\cite{Payez_2015} as a reference.

\begin{figure}[htbp]
    \centering
    \includegraphics[keepaspectratio, scale=0.48]{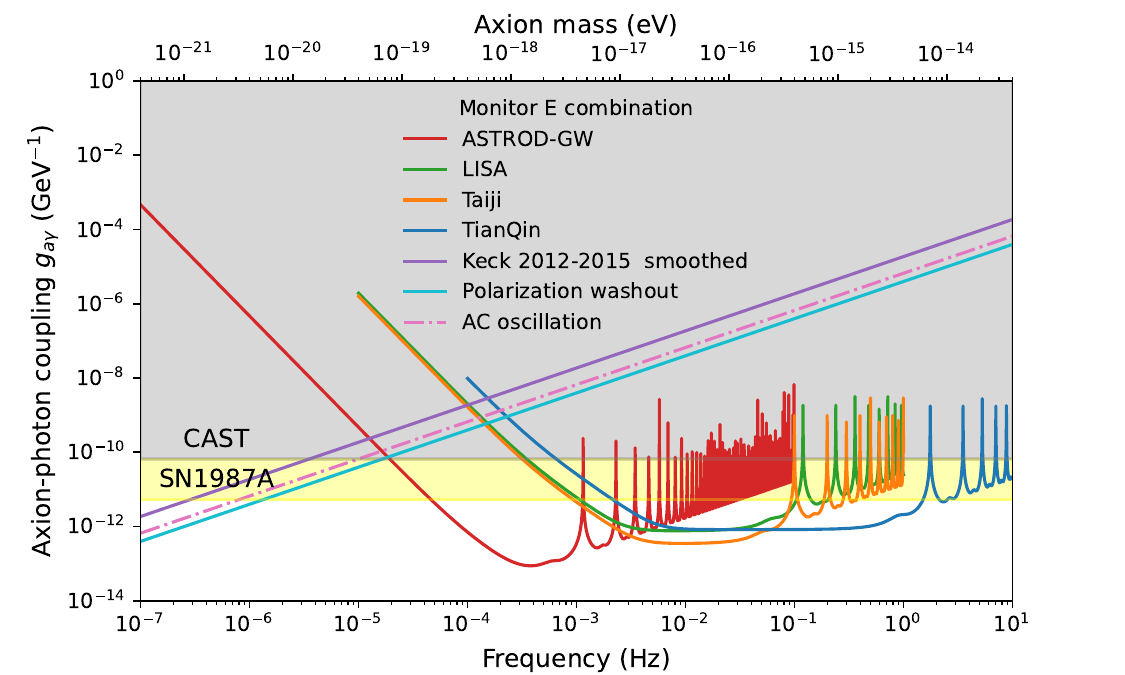}
    \caption{Sensitivity curves of various space-based laser interferometers to the axion-photon coupling $g_{a\gamma}$ under the Monitor-E combination. From left to right, they represent the sensitivities of ASTROD-GW (red), Taiji (orange), LISA (green), and TianQin (blue). The grey area and the yellow area represent the parameter space excluded by the CAST~\cite{2017} and SN1987A~\cite{Payez_2015} experiments, respectively. 
The constraints from CMB polarization measurements are also shown in the plot. The purple line (BICEP/Keck results) represents the improved constraints on axion polarization oscillations in the CMB background. For detailed information, one can refer to~\cite{PhysRevD.105.022006}. 
Based on the Planck results~\cite{PhysRevD.100.015040}, the cyan line and the pink dotted line represent the polarization washout effect and the Alternating Current (AC) oscillation effect with a rotation amplitude of at least $0.1^{\circ}$, respectively. 
The regions above the three diagonal lines mark the excluded parameter space for ultra-light axions.
}
\label{45}
\end{figure}

\begin{figure}[!htbp]
    \centering
    \includegraphics[keepaspectratio, scale=0.48]{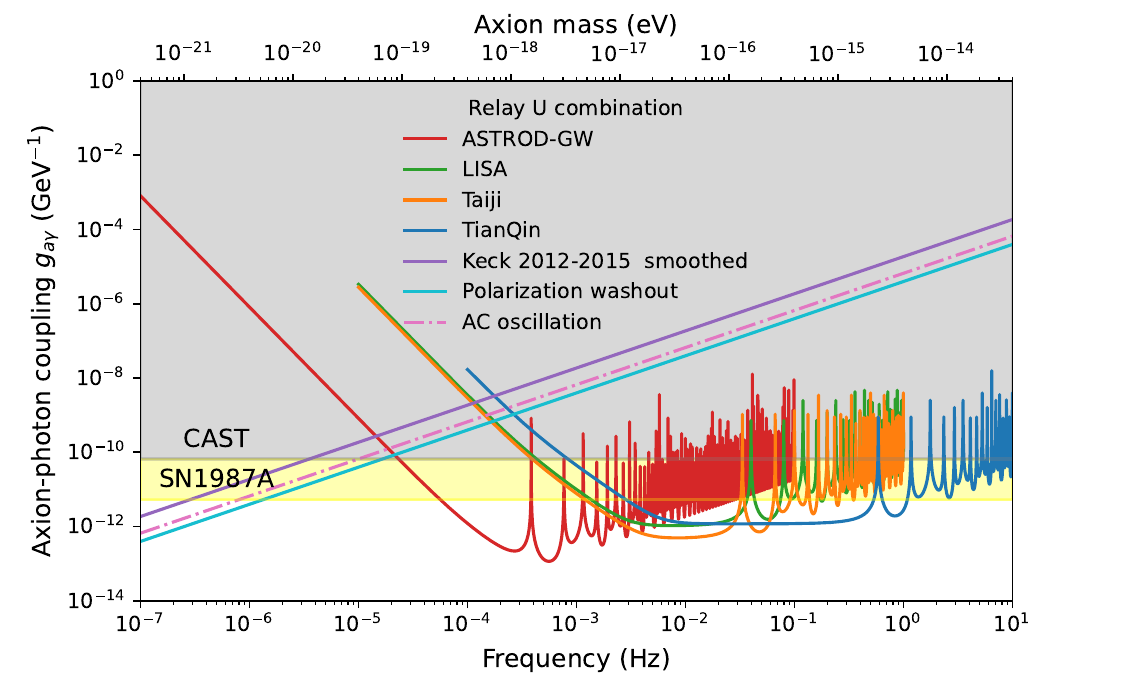}
    \caption{Sensitivity curves of various space-based laser interferometers to the axion-photon coupling $g_{a\gamma}$ under the Relay-U combination. All other captions follow the same conventions as in Fig. \ref{45}.} 
    \label{46}
\end{figure}

\begin{figure}[!htbp]
    \centering
    \includegraphics[keepaspectratio, scale=0.48]{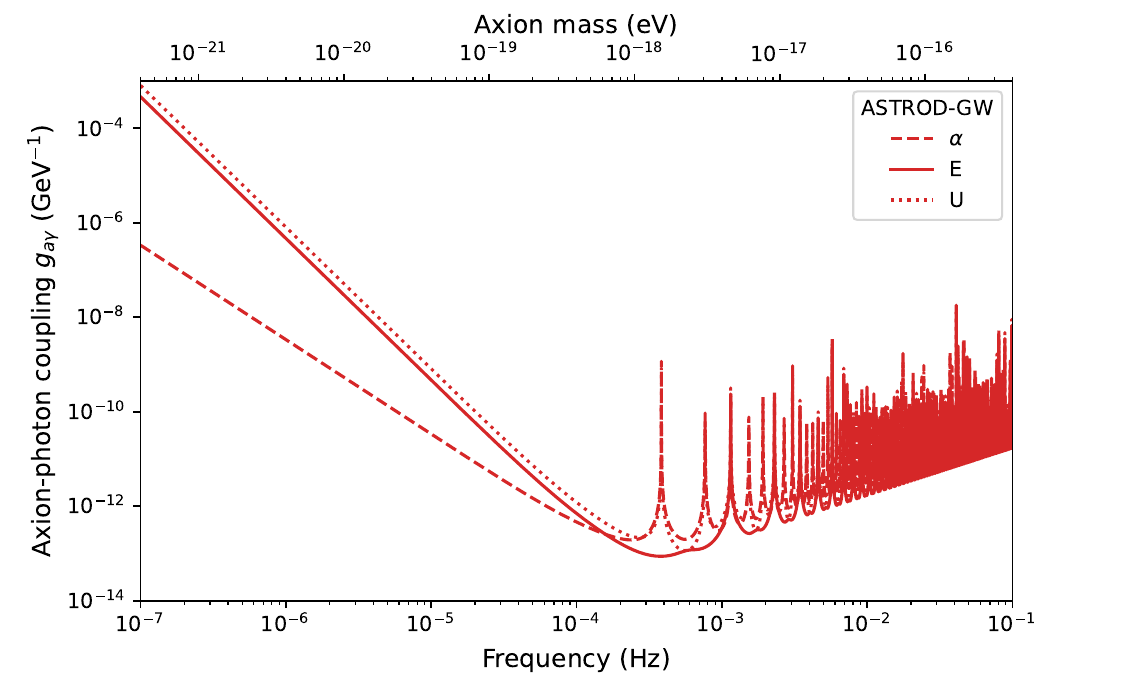}
    \caption{Sensitivity curves of ASTROD-GW to the axion-photon coupling $g_{a\gamma}$ under different TDI combinations. Dashed, solid, and dotted lines denote the sensitivity curves for the Sagnac-$\alpha$, Monitor-E, and Relay-U combinations.}
    \label{41}
\end{figure}

\begin{figure}[htbp]
    \centering
    \includegraphics[keepaspectratio, scale=0.48]{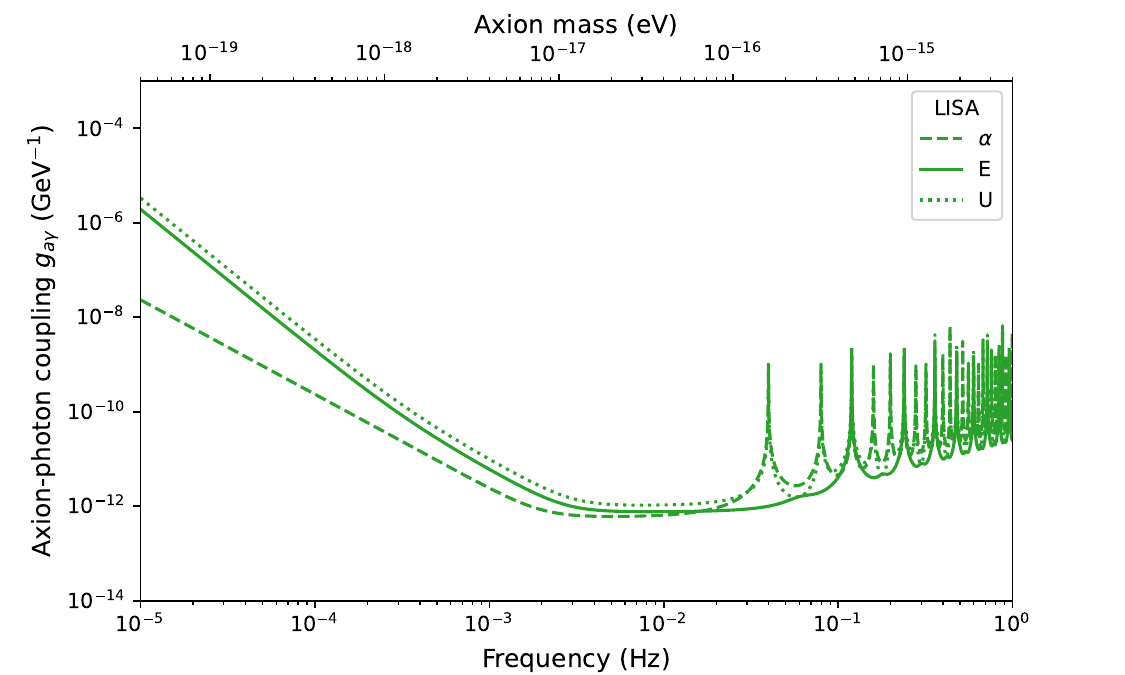}
    \caption{Sensitivity curves of LISA to the axion-photon coupling $g_{a\gamma}$ under different TDI combinations. Dashed, solid, and dotted lines denote the sensitivity curves for the Sagnac-$\alpha$, Monitor-E, and Relay-U combinations.}
    \label{43}
\end{figure}

\begin{figure}[htbp]
    \centering
    \includegraphics[keepaspectratio, scale=0.48]{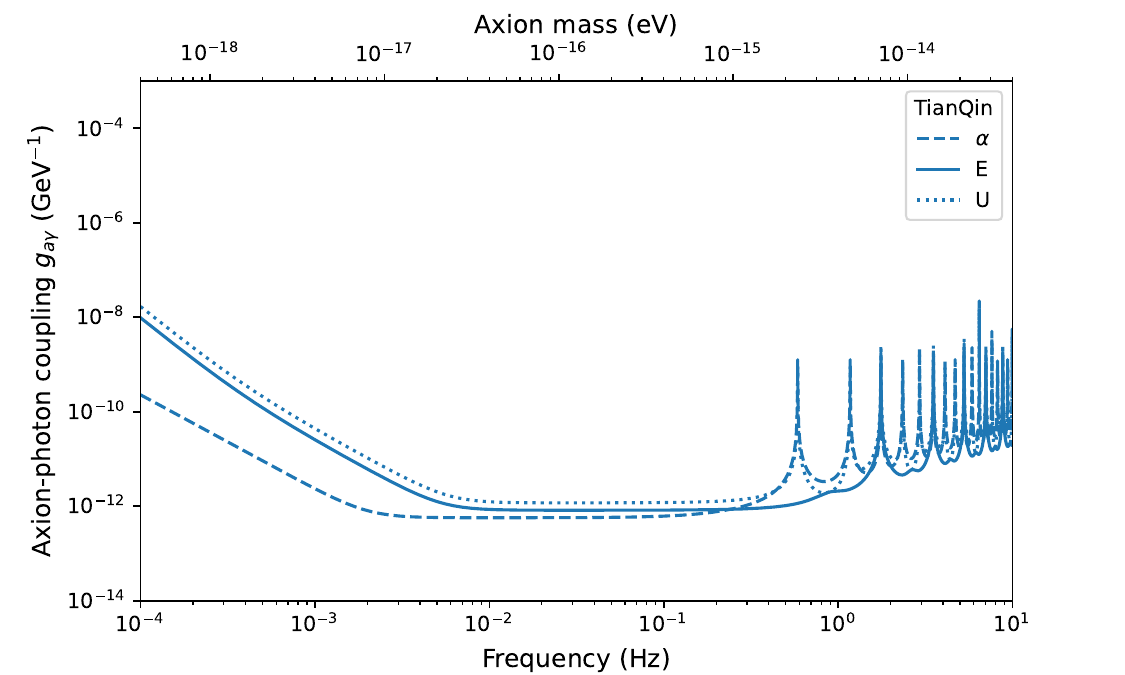}
    \caption{Sensitivity curves of TianQin to the axion-photon coupling $g_{a\gamma}$ under different TDI combinations. Dashed, solid, and dotted lines denote the sensitivity curves for the Sagnac-$\alpha$, Monitor-E, and Relay-U combinations.}
    \label{44}
\end{figure}

\begin{figure}[htbp]
    \centering
    \includegraphics[keepaspectratio, scale=0.48]{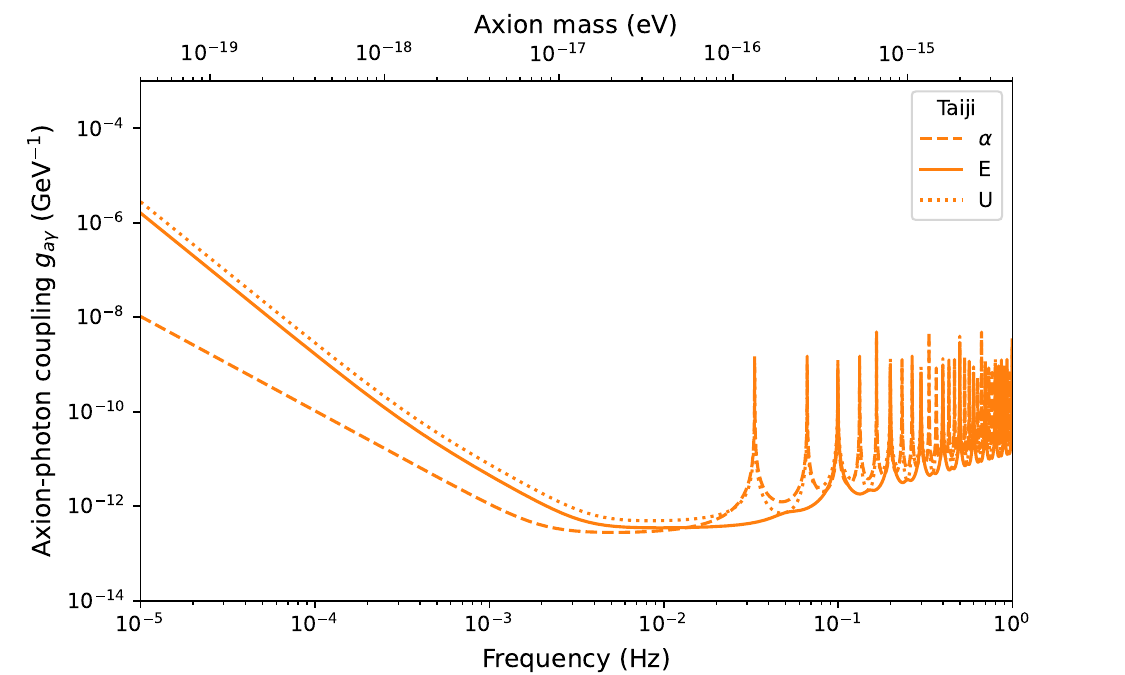}
    \caption{Sensitivity curves of Taiji to the axion-photon coupling $g_{a\gamma}$ under different TDI combinations. Dashed, solid, and dotted lines denote the sensitivity curves for the Sagnac-$\alpha$, Monitor-E, and Relay-U combinations.}
    \label{42}
\end{figure}

The signal-to-noise ratio (SNR) is defined as~\cite{PhysRevX.4.021030}
\begin{equation}\label{eq.21}
    {\rm SNR}=\frac{P_{s}(f)}{P_{n}(f)}.
\end{equation}
The sensitivity is defined as the coupling strength that results in ${\rm SNR}=1$ within a one-year observation period. Substituting Eq.~(\ref{eq.17}), Eq.~(\ref{eq.19}) and Eq.~(\ref{eq.20}) into Eq.~(\ref{eq.21}), we can get the expression for the sensitivity  
\begin{equation}
    g_{a\gamma}=\left(\frac{P_{n}^{C_{i}}\omega^{2}}{\rho_{\text{DM}}T|c_{i}|^{2}}\right)^{1/2}.
\end{equation}
We then calculate the sensitivity of different TDI combinations for various detectors, and show the results in Figs.~\ref{45}--\ref{42} and discuss the results below.

Compared with the Sagnac combination, the Monitor (Beacon) and Relay combinations perform relatively better in the high-frequency range, as shown in Figs.~\ref{41}--\ref{42}. In addition, the flat region at the lowest point of the sensitivity curve of the Monitor (Beacon) combination is wider than that of the Sagnac and Relay combinations, which indicates better detection performance in this frequency range.

Furthermore, for the four types of space-based laser interferometers, their detection ranges are complementary, with each one exhibiting optimal sensitivity within specific frequency bands. As shown in Figs.~\ref{45}--\ref{46} that ASTROD-GW has the best performance within the range $[10^{-7}, 10^{-3}]$\,Hz, LISA and Taiji have the best performance within the range $[10^{-3}, 10^{-1}]$\,Hz, and TianQin has the best performance within the range $[10^{-1}, 10^{1}]$\,Hz. The optimal frequency bands for each detector arise from its arm length and design parameters.


Since the arm length of ASTROD-GW is approximately 100 times longer than those of LISA, Taiji, and TianQin, its theoretical sensitivity is expected to improve by two orders of magnitude. However, as shown in Figs.~\ref{45}--\ref{46}, the sensitivity improvement of ASTROD-GW is limited to about two orders of magnitude in the frequency range of $[10^{-7}, 10^{-3}]$\,Hz. The reason is that, as frequency $f$ increases, the optical metrology system noise $S_{\rm oms}$ dominates. As shown in Fig.~\ref{47}, in the frequency range of $[10^{-3}, 10^{-1}]$\,Hz, $S_{\rm oms}$ governs the sensitivity, while in the frequency range of $[10^{-7}, 10^{-4}]$\,Hz, the acceleration noise $S_{\rm acc}$ dominates. Due to the design of the ASTROD-GW mission, its optical measurement noise is approximately one to two orders of magnitude higher than that of other space-based detectors, while the test mass acceleration noise is about one order of magnitude lower than that of other detectors \cite{NI_2013}. This is the reason why ASTROD-GW has better sensitivity only in the low-frequency range.

The coherence time $\tau_{c}$ and coherence length $\lambda_{c}$ of the axion-like dark matter are
$ \tau_{c}=\frac{2\pi\hbar}{m v_{\text{DM}}^{2}}\sim 1\text{yr}
    \left(\frac{10^{-16}\text{eV}}{m}\right),$
$ \lambda_{c}=\frac{2\pi\hbar}{mv_{\text{DM}}}\sim 10^{4}\text{Gm}\left(\frac{10^{-16}\text{eV}}{m}\right)$,
where $v_{\rm DM}\sim 10^{-3}$ represents the velocity of axion-like dark matter.
When the coherence time and coherence length of the axion-like dark matter are much longer than the observation time $T$ and the detector arm length $L$, repectively, the spatial and temporal dependence of the axion-like dark matter can be neglected. 
For $T=1\,\text{yr}>\tau_{c}$, corresponding to axion-like dark matter with mass greater than $10^{-16}\,\rm eV$, the axion-like field cannot be regarded as a fully coherent field and it should be modeled as the superposition of several random harmonics~\cite{PhysRevD.97.123006}. One data processing method involves dividing the data set into several segments with durations less than $\tau_{c}$, searching for coherent axion dark matter signals within each of these segments. Then, a correction factor $(\frac{T}{\tau_{c}})^{\frac{1}{4}}$ needs to be included in the calculation of the sensitivity curve~\cite{Yao:2024fie, Gu__2025,PhysRevX.4.021030}. 
For $L>\lambda_{c}$, the axion-like field loses spatial coherence. In addition, due to the cancellation effect, the sensitivity curve will exhibit oscillations and have multiple singular points. Therefore, the sensitivity declines sharply in this region and needs to be corrected. We need to limit and truncate the detection range for each detector to ensure that the axion-like dark matter maintains coherence within its respective detection range.

\color{black}
\section{Conclusion and Discussion}
\label{Conclusion}
In this work, we introduce a class of axion-like dark matter detection schemes based on space-based gravitational wave detectors. By converting the laser beam between spacecrafts from linear polarization to circular polarization, we utilize the axion-induced birefringence effect for the detection of axion-like dark matter. For example, we primarily evaluate the sensitivity of the ASTROD-GW mission to  the axion-photon coupling in the $\mu$\,Hz frequency band. We find that detectors with different arm lengths have their optimal sensitivity curves in different frequency bands, as shown in Fig.~\ref{45}. The optimal detection frequency window of ASTROD-GW is $[10^{-7}, 10^{-3}]$\,Hz, corresponding to an axion mass range of approximately $10^{-20}$ to $10^{-17}$ eV. Therefore, ASTROD-GW has the potential to expand the detectable mass range of axion-like dark matter into the lower-mass regime.

{

One important issue involves distinguishing between gravitational wave (GW) signals and axion-like signals. For the conventional scheme in space-based gravitational wave detectors, multiple approaches have been proposed to discriminate ultralight dark matter signals from GW signals, including the utilization of different TDI channels~\cite{Xu:2025rfv}, accounting for the characteristic frequency modulation of ultralight dark matter~\cite{Yao:2025vgy}, and employing Bayesian analysis methods~\cite{Gue:2025iab}. These techniques can be suitably adapted to our case. In principle, the time-frequency characteristics of axion-like dark matter are different from those of monochromatic GW signals, which makes their discrimination viable.

It is worth noting that our optical system design is purely theoretical. For the practical design of optical paths, one can refer to~\cite{Gu__2025}, which accounts for light reflection effects and proposes that one of the light components retains linear polarization. A straightforward analysis shows that the phase difference obtained using two circularly polarized light paths is twice as large as that obtained from one circularly polarized light path and one linearly polarized light path. As a result, the sensitivity may be reduced by a factor of two. 
However, the scheme in~\cite{Gu__2025} offers higher reliability, although its specific implementation requires further investigation. By retaining the linearly polarized light path so that the axion signal does not respond in this path, while suppressing the gravitational wave signal in the other circularly polarized light path, in principle the axion signal can be isolated from the gravitational wave background.

}

In addition, we conduct a comparative analysis of three TDI combinations: Monitor, Beacon, and Relay. We find that Monitor and Beacon combinations have better sensitivity in the high-frequency range, compared with the Sagnac combination. Moreover, their optimal sensitivity range is broader than that of the Sagnac combination. In contrast, their performance is relatively limited in the low-frequency range. These results indicate that TDI combinations exhibit frequency-dependent performance variations.
In future work, we will investigate other TDI combinations, with particular focus on TDI 2.0 configurations, and evaluate their optimal performance across different frequency bands.

\begin{acknowledgments}
This work is supported by 
the National Key Research and Development Program of China (No. 2023YFC2206200, No. 2021YFC2201901), 
the National Natural Science Foundation of China (No. 12375059),  and the Project of National Astronomical Observatories, Chinese Academy of Sciences (No. E4TG6601). We thank Run-Min Yao and Gang Wang for many helpful discussions.

\end{acknowledgments}

\appendix


\section{Sagnac combination}


The Sagnac-$\alpha$ combination is given by~\cite{Tinto:2020fcc}  
\begin{align}
    S_{\alpha}(t)=&\ \Phi_{13}(t)+D_{1}\Phi_{32}(t)+D_{13}\Phi_{21}(t)\nonumber\\
    &-\Phi_{12}(t)-D_{1'}\Phi_{23}(t)-D_{1'2'}\Phi_{31}(t),
\end{align}
where $D_{kl}\Phi_{ij}(t)\equiv\Phi_{ij}(t-L_{k}-L_{l})$. Here, $\Phi_{ij}(t)$ represents the phase measurement data at time $t$. The first index refers to the spacecraft that transmits the signal, the second indicator refers to the spacecraft that receives the signal. The concrete form is given by~\cite{PhysRevD.65.022004}
\begin{align}
    \Phi_{ij}(t)=&\ n_{i}(t)-n_{j}(t-L_{i})+\psi_{ij}(t)\nonumber\\
    &+s_{ij}^{\rm oms}(t) -\left[ {s}_{ij}^{  \rm acc}(t)-{s}_{ji}^{ \rm acc}(t-L_{i})\right].
\end{align}
The time-varying part of the phase includes contributions from the laser phase noise $n_{i}(t)$,  optical metrology system noise $s^{\rm oms}(t)$, test mass acceleration noise ${s}^{\,\rm acc}(t)$, and the scientific signal of interest $\psi_{ij}(t)$. In this work, this signal is related to the relative frequency fluctuation signal $\eta_{i}(t)$ in Eq.~\eqref{6}, which is caused by the axion-induced birefringence effect. For simplicity, we set $L_{i}=L$ and assume that the laser frequency noise has been eliminated. By performing the Sagnac-$\alpha$ combination on the phase measurement, we obtain 
\begin{align}
    S_{\alpha}(t)=&\ \Phi_{13}(t)+\Phi_{32}(t-L)+\Phi_{21}(t-2L)\nonumber\\
    &-\Phi_{12}(t)-\Phi_{23}(t-L)-\Phi_{31}(t-2L)\nonumber\\
    =&\ \alpha(t)+n_{\rm oms}(t)+n_{\rm acc}(t).
\end{align}

The signal part $\alpha (t)$ is given by
\begin{equation}
\alpha(t)=\eta_{1}+D_{1}\eta_{3}+D_{13}\eta_{2}-\eta_{1'}-D_{1'}\eta_{2'}-D_{1'2'}\eta_{3'} .  
\end{equation}
By substituting the data streams from Eqs.~\eqref{6}-\eqref{7} into $\alpha(t)$, we obtain the time-domain data
\begin{equation}
    \alpha(t)=\frac{img_{a\gamma}}{\omega}a_{0}e^{i(mt+\theta_{0})}(1-e^{-i3mL}).
\end{equation}
The one-sided PSD of the signal is defined as
\begin{equation}
    P_{s}^{\alpha}(f)=2\frac{|\tilde{\alpha}(f)|^{2}}{T}=\frac{4\rho_{\rm  DM}g_{a\gamma}^{2}T}{\omega^{2}}\sin^{2}(3\pi fL),
\end{equation}
where $T$ is the observation period, the frequency $f$ is related to the mass of axion-like particles via $m=2\pi f$.

To evaluate the one-sided PSD of the noise, we calculate the frequency domain expressions of the two types of noise separately and linearly combine the noise lines of each spacecraft vertex. The time delay is manifested as a delay operator $D_{i}$ in the frequency domain. Since we assume constant arm length, the subscript for $D_{i}$ is omitted. The frequency domain expression of the optical metrology system noise is given by
\begin{align}
    n_{\rm oms}(f)=&\ s_{13}^{\rm oms}(f)+Ds_{32}^{\rm oms}(f)+D^{2}s_{21}^{\rm oms}(f)\nonumber\\
    &- s_{12}^{\rm oms}(f) -Ds_{23}^{\rm oms}(f) -D^{2}s_{31}^{\rm oms}(f),
\end{align}
and the frequency domain expression of test mass acceleration noise is
\begin{align}
    n_{\rm acc}(f)=&\ (D^3-1)s_{13}^{\rm acc}(f)+(D^{2}-D)s_{31}^{\rm acc}(f)\nonumber\\
    &+(D^{2}-D)s_{32}^{\rm acc}(f)+(D-D^{2})s_{23}^{\rm acc}(f)\nonumber\\
    &+(D-D^{2})s_{21}^{\rm acc}(f)+(1-D^3)s_{12}^{\rm acc}(f).
\end{align}

By assuming that each optical assembly has the same noise spectrum, and that the optical metrology system noise $n_{\rm oms}(f)$ and the test mass acceleration noise $n_{\rm acc}(f)$ are uncorrelated with each other, the one-sided PSD of the noise is calculated as 
\begin{align}
    P_{n}^{\alpha}&=\left \langle n(f)n^{*}(f)\right \rangle\simeq \left \langle n_{\rm oms}(f)n^{*}_{\rm oms}(f)\right \rangle+\left \langle n_{\rm acc}(f)n^{*}_{\rm acc}(f)\right \rangle\nonumber\\
    &=6S_{\rm oms}(f)+8\left[\sin^2(3f/2f_*)+2\sin^2\left({f}/{2f_*}\right)\right]S_{\rm acc}(f),
\end{align}
where $f_{*}=c/(2\pi L)$ is defined as the characteristic frequency scale of the detector.
\color{black}

\bibliography{ref.bib}

 \end{document}